\documentclass[12pt]{article}
\usepackage{amsfonts}
\usepackage{amsmath}
\usepackage{graphicx}
\usepackage[T1]{fontenc}

\newtheorem{theorem}{Theorem}
\newtheorem{definition}{Definition}
\newtheorem{lemma}{Lemma}
\newtheorem{proposition}{Proposition}

\newtheorem{remark}{Remark}

\def\id{{\bf 1}\!\!{\rm I}}

\begin{document}

\centerline{\bf \large A Forward Quantum Markov Field on Graphs    }


\begin{center}
{\sc Abdessatar Souissi}\\
\textit{$^1$ Department of Preparatory program, College of Business Administration,
Qassim University,P.O.BOX:6666- Buraidah:51452, Saudi Arabia.\\
 $^2$ Preparatory Institute for Scientific and Technical Studies, Carthage University,
   Amilcar1054, Tunisia.}\\
 E-mail:{\, \sf a.souaissi@qu.edu.sa, \, \,   \ abdessattar.souissi@ipest.rnu.tn}
\end{center}

\vspace{1cm}

\begin{abstract}
In this paper, we propose a class of quantum Markov fields QMF on a graphs $G= (V,E)$.  The Markov structure  
of the considered QMF is  investigated  in the finer structure of a quasi-local algebrav $\mathcal{A}_V$ of observables based over a graphs $G$.
 Namely,  the  considered Markovian fields are infinite volume states defined through  a generating  couple
  $(\varphi^{(0)}, (\mathcal{E}_{\{y\}\cup N_y}))$ of a product state $\varphi^{(0)}$ on $\mathcal{A}_V$ and a
   family of local transition expectations $\mathcal{E}_{\{y\}\cup N_y}$ based on a vertex $y$ and the set of it nearest-neighbors.
 The main result of the paper concerns the existence and the uniqueness of QMF associated with a couple $(\varphi^{(0)}, (\mathcal{E}_{\{y\}\cup N_y}))$
   for  on an important class of graphs including trees strictly. 
\end{abstract}


 \section{Introduction}
Markov random fields  \cite{D}, \cite{Spi}, \cite{Za} have become a standard tools 
in several areas such as classical probability, statistical physics,  computer science, image segmentation. 
 However, a satisfactory theory of quantum  Markov fields is still missing. 

The first attempts to construct such a theory are \cite{Acc.Fid}, \cite{Acc.Fi.entangled}, \cite{AMSo}.
 These papers extended the Dobrushin-Markov random fields on the integer lattice $\mathbf Z^\nu$ ( see
  \cite{D}, \cite{D1})  to a quantum setting.
In \cite{MBSo1}, \cite{AMSa1}, \cite{AMSa2} and \cite{AMSa3} quantum Markov chains on the Cayley  were constructed and phase transitions were investigated through Ising type models.
In \cite{MBSo1} some algebraic properties of the disordered phase associated with an
Ising type model were studied on the same kind of trees.\\

From a quantum probabilistic viewpoint QMF are multi-dimensional extension of one-dimensional quantum Markov chains and states  introduced  by L. Accardi in  
\cite{Acc74}.
In \cite{AOM.2010} the notion of generalized quantum Markov states has been extended
to fields, i.e. to quasi-local algebras over  graphs possesing a hierarchy property.
Namely, the notions of \textit{d-Markov chains} and \textit{generalized Markov states} on graphs
have been investigated, as natural extension to graphs of one-dimensional quantum Markov
chains.   
 
 The non-commutative Markov property  \cite{[Acc75]} play a crucial role in the description of quantum Markovian states. 
In \cite{AMSa1}-\cite{AMSa3} a multi-dimensional  Markov property was investigated on the fine structure of the Cayley trees
 and in \cite{MBSo1}, \cite{MBSo2} additional interactions between  one-level nearest neighbors vertices
  on the same kind of trees were investigated.  
  
In \cite{AMSo} a backward quantum Markov fields was constructed on a tensor algebra over
arbitrary graph. Namely,  the noncommutative Markov property on the finer structure  of the  considered graph 
was investigated across conditional expectations in the sense of \cite{Acc.Cecc82} and \cite{Acc.local79}.\\

In this paper, we investigate a tessellation on an infinite  graph $G = (V, E)$ that splits the vertex set into two infinite subsets: 
$V_{\infty}$ and its complementary. The quantum  Markov prperty is investigated  for local structure of a UHF $\mathcal{A}_V$ of observables over the graph $G$. Namely, 
 a non-commutative extension of the notion of Markov triplet studied in \cite{Spa} is carried out for transition expectation. 
  The construction of the forward Markov field is based on a reference product state $\varphi^{(0)}= \bigotimes_{x\in V}\varphi_x$ together with a
   famely of transition expectation $\{ \mathcal{E}_y , y\in {V}_\infty\}$ acting on the observables localized on the local algebra over the site $y$ and the 
   its nearest-neighbors. 
   Mainly, we show the existence and the uniqueness of a forward quantum Markov field associated with the couple 
   $(\varphi^{(0)}, \{ \mathcal{E}_y , y\in\mathcal{V}_\infty\})$. The noncommutative nature of the considered transition 
   expectations requires an enumeration,  on the set $V_{\infty}$ on which the resulting quantum Markov field is strongly attached.
   The provided QMF are of great interest  on tree-like graphs.
   We stress that a concrete Ising and XY-type models in connections will be the purpose of a paper in preparation aiming to apply the results of the present paper
   to the study of  phenomena  of phase transitions for several  graphs.

Let us briefly mention the organization of the paper,
after preliminary informations in Section \ref{prel}. We construct, in section \ref{section_tess}, a tessellation on an
infinite, locally finite, connected undirected graph. 
In section \ref{sec_Mp}, we investigate the quantum Markov property on fine structure of the graph for  local transition expectations w.r.t.  precise Markov triplets. 
Section \ref{section_main} is devoted to the definition of forward  quantum Markov
fiels and the  formulation of the main result which concerns the existence and the
uniqueness of quantum Markov field for a large class of graphs including the tree.
Section \ref{proof_main} is devoted to prove the main results. In section \ref{annex} we prove some auxiliary results on graphs.

\section{Preliminaries}\label{prel}

Let $G=(V, E)$ be a  ( non-oriented  simple ) graph, where $V$ is a nonempty set  and $E$
 is identified as a subset of non-ordered pairs
of $V$, i.e. $$E\subset \{ \{x,y \} \mid x,y \in E, x\ne y\}\setminus \{(x,x), \; x\in E\}.$$
Elements of $V$ and $E$ are called, respectively, \textit{vertices}
and \textit{edges}. Two vertices $x$ and $y$ are said to be
\textit{nearest neighbors}, written $x\sim y$, if and only if $\{x,y\}\in E$.\\
 For $y\in V$ we denote its nearest neighbors by
\begin{equation}\label{Ny}
N_y := \{x\in V \mid y\sim x\}
\end{equation}
Notice that $x\notin N_x$. The set $\{y\}\cup N_y$ is called
\textit{interaction domain} or \textit{plaquette} at $y$.  In the sequel, the graph $G$  is assumed to be locally finite in the sense that $|N_x|<\infty$ for each  $x\in V$. An \textit{edge path }  joining two vertices $x$ and $y$ is a finite sequence of edges
$x=x_0\sim x_1\sim\dots x_{d-1}\sim x_d=y$, in this case $d$ is the length of the edge path. The graph is said to be \textit{connected }
if every two disjoint vertices can be joined by an edge path. In the sequel, we assume that the graph $G$ is infinite, connected and
locally finite. Thus the set $V$ is automatically countable.
For  $\Lambda\subset V$ nonempty, define
\begin{itemize}
\item \textit{Internal boundary}:
\begin{equation}\label{boundary}
\overleftarrow{\partial} \Lambda := \{x\in \Lambda \; \mid \; \exists y\in \Lambda^c;\quad  x\sim y\}
\end{equation}
\item\textit{Interior}:
\begin{equation}\label{interior}
\overset{\circ}{\Lambda} := \Lambda\setminus \overleftarrow\partial \Lambda
\end{equation}
\item \textit{External boundary}:
\begin{equation}\label{external}
\overrightarrow\partial \Lambda := \{y\in \Lambda^c \; \mid \quad \exists x\in
\Lambda;\quad  x\sim y\}
\end{equation}
\item \textit{External closure}:
\begin{equation}\label{closure}
\overline{\Lambda} := \Lambda\cup \overrightarrow\partial \Lambda
\end{equation}
\end{itemize}
 Denote $F$ the set of all finite subsets of the vertex $V$ and $\mathcal F$  the net generated by $F$ ordered by  the inclusion $"\subset"$.
To each site $x\in V$, we associate a finite dimensional $C^\ast$-algebra of observable $\mathcal A_x$. Put $\mathcal A_{\Lambda} = \bigotimes_{x\in \Lambda}\mathcal A_x$ the algebra of observables localized in
a finite region $\Lambda\subset V$, where $\otimes$ denotes the algebraic tensor product.\\
Denote  $\mathcal A $  the quasi-local algebra (see section 2.6 of \cite{BR}))
obtained by the C$^\ast$-inductive limit associated to the directed system of algebras
$\{\mathcal A_{\Lambda}\}_{\Lambda\in \mathcal F}$ with the  embedding
\begin{equation}
j_{\Lambda, \tilde\Lambda}: a_\Lambda\in\mathcal A_{\Lambda}\mapsto  a_{\Lambda}\otimes \id_{\tilde \Lambda\setminus \Lambda}\in\mathcal A_{\tilde\Lambda}
\end{equation}
In particular, if $\tilde\Lambda = V$ one gets the following identification
 \begin{equation}
 \mathcal A_\Lambda \cong j_\Lambda(\mathcal A_\Lambda) = \mathcal A_\Lambda\otimes \id_{\Lambda^c}
 \end{equation}
  this leads to  the local algebra
 \begin{equation}
 \mathcal A_{loc} = \bigcup_{\Lambda\in  F}\mathcal A_{\Lambda}
 \end{equation}
  The quasi-local algebra $\mathcal A$ is then the closure of $\mathcal A_{loc}$
 $$\mathcal A= \overline{A_{loc}} = \overline{ \bigcup_{\Lambda\in  F}\mathcal A_{\Lambda}}.$$
 The natural embedding of $\mathcal A_x$ into $\mathcal A_V$ will be denoted by
 $$j_x: a\in\mathcal A_x \mapsto a\otimes \id_{\{x\}^c}.$$
 Similarly, for $\Lambda \in \mathcal F$ the embedding $j_\Lambda$ can be written as follows
$$j_\Lambda = \bigotimes_{x\in\Lambda}j_x.$$

 \section{Tessellations on graphs}\label{section_tess}
Let $G = (V, E)$ be an infinite connected graph which is locally finite (i.e. for each $x\in V$ the set $N_x$ if its nearest neighbors  is finite.)\\
 Fix a "root"  $y_1 \in V$ and define by induction the following sets:
\begin{equation}
   V_{0,1}:=\{ y_1\};\quad  V_1 = \{o\}\cup N_{o}\label{root}
\end{equation}
 Having defined $V_{0,n}$, put
\begin{equation}\label{Vn}
V_n := \overline{V}_{0,n} = \bigcup_{y \in V_{0,n}} \left( \{y \} \cup N_y \right)
\end{equation}
\begin{equation}\label{v0,n+1}
 V_{0,n+1}:=V_{0,n}\cup\overrightarrow{\partial} V_n
\end{equation}


Put
\begin{equation}\label{Vinfty}
V_{\infty} := \bigcup_{n\geq 1} V_{0,n}
\end{equation}
Each element of $V\setminus V_{\infty}$ belongs to the plaquette at a certain element
of $V_{\infty}$
$$V\setminus V_{\infty} = \bigcup_{y\in V_{\infty}}\overrightarrow\partial\{y\}$$

In the sequel we assume that the graph $G$ becomes totally disconnected by removing $V_{\infty}$ or $V\setminus V_{\infty}$. This means that every edge has an endpoint in $V_{\infty}$ and an endpoint in $V\setminus V_{\infty}$. Note that this property is satisfied by an important of graphs such as trees and the multi-dimensional integer lattice  $\mathbb Z^d$.
\begin{proposition}\label{prop_V_n}
For each $n\in \mathbf N^*$, let $E_n = \{\{x,y\}\in E; \quad x,y\in V_n\}$  then the following assertions holds
\begin{description}
\item[(i)] The subgraph  $G_n:=(V_n, E_n)$ is finite and connected,
\item[(ii)] $V = \bigcup_{n\geq 1}V_n$ and $E = \bigcup_{n}E_n$.
\end{description}
\end{proposition}

It follows that the sequence $(V_n)_{n\in\mathbb N}$ is exhaustive for the vertex set $V$. i.e. $V_n\subset V_{n+1}$  for any finite subset $\Lambda$ of $V$
there exists an integer $n$ such that $\Lambda \subseteq V_n$.
 One gets
\begin{equation}
V_{\infty} := \bigcup_{n\geq 0} \overrightarrow\partial V_{n}\label{Vee0}
\end{equation}
and
\begin{equation}
V=\bigcup_{y\in V_{\infty}}\{y\}\cup N_y\label{VinftyV}
\end{equation}
In what follows, the  family $\{V_{0,n}; \quad n=1,2,\cdots\}$ ) will be called a \textit{tessellation}
of the graph $G$.
For each $n\in\mathbf N $and $y\in \overrightarrow\partial V_n$, put
\begin{equation}\label{Nysp}
 N^{(p)}_{y}:= N_y\cap\overleftarrow{\partial}V_n \, \, \, N^{(s)}_{y}:=N_y\cap \overleftarrow{\partial} V_{n+1},\, \,
  \, N^{(0)}_{y}:= N_y\setminus(N_y^{(p)}\cup N_y^{(s)})
\end{equation}
 here $N^{(s)}_{y}$ and $N^{(p)}_{y}$ refer respectively to the sets of \textit{the direct successors vertices} and \textit{ direct previous vertices} of the vertex $y$.

\begin{proposition}\label{partition}
For each integer $n\geq 1$, the following assertions holds
\begin{enumerate}
\item[(i)]$\overleftarrow{\partial}V_{n+1}  =   \bigcup_{y\in \overrightarrow\partial V_n }N^{(s)}_{y},$
\item[(ii)] $\overleftarrow{\partial}V_n  = \bigcup_{y\in \overrightarrow\partial V_n}N^{(p)}_{y}.$
\end{enumerate}
\end{proposition}

   \begin{remark}
 Note that generally the inclusion  in (i) of Proposition \ref{partition}  is strict.
 In fact the following case may occur: $x\in V_{n+1}\setminus \overline V_{n}$ with $N_x =\{y\}$ for some $y\in\overrightarrow\partial V_n$.
\end{remark}
For the sake of simplicity, we assume, from now on, that the  graph $G= (V,E)$  satisfy the following conditions:
\begin{equation}\label{N0=0}
 N_y^{(0)} = \emptyset
\end{equation}

 \begin{equation}\label{(s)-property}
N_y^{(s)}\cap N_z^{(s)}=\emptyset
\end{equation}
 for every integer $n$ and   every $y,z\in \overrightarrow{\partial} V_n$ with $y\ne z$.

The properties (\ref{N0=0})  and (\ref{(s)-property}) are satisfied by an important class of graphs including trees (which are connected graphs with no cycles).
 Indeed, the property  (\ref{(s)-property}) is satisfied by any tree-like graph and  (\ref{partial(p)=partial}) is satisfied for trees  without  degree-one vertices and
  in the case of  the  integer lattices $\mathbb Z^d$.
\begin{proposition}\label{tree_S_property}
 Every tree enjoys the properties (\ref{N0=0})  and (\ref{(s)-property})  .
\end{proposition}

From Proposition \ref{V_n} for each integer $n$ the subset  $V_n$ given by (\ref{Vn}) is
finite then its external boundary $\overrightarrow\partial V_n$ is also finite.
The following  enumeration arises naturally.\\
 Let us fix an  enumeration of the set $\overrightarrow\partial V_n$
 
\begin{equation}\label{enumeration_vec_vn}
\overrightarrow\partial V_n = \left\{ y_{1}^{(n)}, \dots, y_{|\overrightarrow\partial V_n|}^{(n)}\right\}
\end{equation}
According to (\ref{Vee0}), this leads  to an enumeration on the  set $V_{\infty}$ given by (\ref{Vinfty}).

\section{Quantum transition expectations and Markov triples}\label{sec_Mp}

Recall that, a Umegaki conditional expectation is a norm one projection  $E$ from a C$^{\ast}$-algebra $\mathcal A$ into a C$^\ast$-subalgebra $\mathcal B$.
\begin{definition}
  Let $\mathcal A$ and $\mathcal B$ be a C$^\ast$-subalgebra. A completely positive identity preserving  map $\mathcal E: \mathcal A \to \mathcal B$  is called transition expectation.\\
  If $\mathcal C$ is a given   C$^\ast$-subalgebra of $\mathcal A$ such that
  \begin{equation}\label{Markov_property}
\mathcal E(\mathcal C'\cap \mathcal A)\subseteq\mathcal C'\cap \mathcal B
\end{equation}
the map   $\mathcal E$ is called \textbf{ Markov transition expectation (MTE)}  with respect to the triplet $(\mathcal A, \mathcal B, \mathcal C).$  Such a  triplet will be referred
as \textbf{Quantum Markov triplet} for the transition expectation $\mathcal E$.
\end{definition}

Note that, the Markov property (\ref{Markov_property}) was  first formulated  in \cite{Acc74}. The notion of quantum triplet Markov consists a noncommutative extension of classical Markov triple studied in \cite{Spa}.

Let $\Lambda_1, \Lambda_2\in\mathcal F,$ with $\Lambda_2\subseteq \Lambda_2$. Let $\mathcal E_{\Lambda_1,\Lambda_2}$ be a Markov transition expectation
 w.r.t the triplet  $(\mathcal A_{ V}, \mathcal A_{\Lambda_1}, \mathcal A_{\Lambda_2})$. Since $\mathcal A_{\Lambda_2 }'=\mathcal A_{\Lambda_2^c}$
  the Markov property  (\ref{Markov_property}) becomes
\begin{equation}\label{Markov_prop_Lab12}
 \mathcal E_{\Lambda_1,\Lambda_2}(\mathcal A_{\Lambda_2^c }) \subseteq \mathcal A_{ \Lambda_1\setminus\Lambda_2}
\end{equation}

Now in order to investigate the Markov property on the finer  structure of  considered graph we consider for each $y\in V_{\infty}$ a Markov transition expectation
$\mathcal E_{y}$ w.r.t. the triplet $(\mathcal{A}_{\{y\}\cup N_{y}}, \mathcal A_{N_y^{(p)}} , \mathcal A_{N_y^{(s)}})$. This leads to
  \begin{equation}\label{MP_Plaquette}
  \mathcal{E}_{y}(\mathcal{A}_{\{y\}\cup N_y })\subseteq \mathcal A_{N_y^{(s)}}.
  \end{equation}

 \begin{lemma}\label{lemma_MTE_level}
Let $n$ be an integer and $\vec\partial V_n$ be enumerated as in (\ref{enumeration_vec_vn}). If for each $j=1,\dots, |\vec\partial V_n|$ a MTE
 $\mathcal E_{y_{j}^{(n)}}$  w.r.t. the triplet $(\mathcal{A}_{\{y_{j}^{(n)}\}\cup N_{y_{j}^{(n)}}}, \mathcal{A}_{N_{y_{j}^{(n)}}^{(s)}},
 \mathcal{A}_{N_{y_{j}^{(n)}}^{(p)}})$
 then the map
 \begin{equation}\label{E_vecpartialV_n}
 \mathcal E_{n, n+1}:= \mathcal{E}_{y_{ |\overrightarrow\partial V_n|}^{(n)}}\circ\cdots \circ\mathcal{E}_{y_{1}^{(n)}}
 \end{equation}
 is a Markov transition expectation w.r.t. the triplet
  $\left( \mathcal{A}_{ V_{n+1} \setminus \overset{\circ}{V}_{n}},
  \mathcal{A}_{V_{n+1} \setminus\overline{V}_{n}}, \mathcal{A}_{\overline{V}_{n}}\right).$

 In particular,  the equation
 \begin{equation}\label{Mplevel}
 \mathcal{E}_{n, n+1}\left( \mathcal{A}_{ V_{n+1} \setminus \overset{\circ}{V}_{n}} \right)\subseteq \mathcal{A}_{V_{n+1} \setminus\overline{V}_{n}}
 \end{equation}
 will be referred as the  level Markov property.
\end{lemma}
  \textbf{Proof}. Since each $\mathcal{E}_{y_{j}^{(n)}}$ is completely positive and identity preserving then, from (\ref{E_vecpartialV_n}),
  the map $\mathcal{E}_{n,n+1}$ is completely positive identity preserving.
    By Proposition \ref{partition} and (\ref{N0=0}) one has
    $$
    \bigcup_{y\in \overrightarrow{\partial} V_n}\{y\}\cup N_y =
    \overleftarrow{\partial}V_n\cup\overrightarrow\partial V_n\cup\overleftarrow\partial V_{n+1}
    =   V_{n+1}\setminus \overset{\circ}{V}_n.
    $$
     According to (\ref{N0=0}) and (\ref{(s)-property}) for $y,z\in \vec\partial V_n$ with $x\ne y$, one gets
    $$
    (\{y\}\cup N_y)\cup (\{z\} N_z) = \left(\{y\}\cup N_y\right)\cup\left(  \{z\}\cup N_z \setminus \left(N^{(p)}_y\cap N^{(p)}_z\right) \right).
    $$

Then, one finds
     \begin{equation}\label{EyEz}\mathcal E_z\circ \mathcal E_y\left(\mathcal A_{\{y\}\cup N_y}\vee \mathcal A_{\{z\}\cup N_z }\right)
     = \mathcal E_z\left(  \mathcal E_y\left(\mathcal A_{\{y\}\cup N_y}\right)\otimes \mathcal{A}_{ \{z\}\cup N_z\setminus(N^{(p)}_y\cap N^{(p)}_z)}\right)
    \end{equation}
    $$
     = \mathcal E_y\left(\mathcal A_{\{y\}\cup N_y}\right)\otimes\mathcal E_z\left( \mathcal{A}_{ \{z\}\cup N_z \setminus  (N^{(p)}_y\cap N^{(p)}_z)}\right)
     \subseteq \mathcal{A}_{N_y^{(s)}}\otimes \mathcal{A}_{N_z^{(s)}}.$$
 Iterating (\ref{EyEz})  on elements of $\{y_1^{(n)}, \cdots, y_{|\vec\partial \Lambda_b|}^{(n)}\}$ , one see that
       $\mathcal{E}_{n,n+1}$ maps $\mathcal{A}_{ \overline{V}_{n+1} \setminus \overset{\circ}{V}_{n}}$
        into $\mathcal{A}_{\overrightarrow V_{n+1}} = \mathcal{A}_{ \overline{V}_{n+1} \setminus V_{n+1}}
        = \mathcal{A}_{ \overline{V}_{n+1} \setminus V_{n+1}}\cap \mathcal A_{V_n}^{'}$.\\
        Therefore, $\mathcal{E}_{n,n+1}$ is a transition expectation w.r.t. the given triplet and it satisfies the Markov property (\ref{Mplevel}).
  $\square$

 \begin{theorem}\label{thm_Evn+1c} Let $\mathcal{E}_{n,n+1}$ be given by (\ref{E_vecpartialV_n}). Then
\begin{equation}\label{E_Vn+1c}
\mathcal E_{V_{n+1}^c} := \mathcal{E}_{n,n+1}\circ \cdots \circ \mathcal E_{0,1}
\end{equation}
is a MTE w.r.t. the triplet $(\mathcal{A}_V, \mathcal{A}_{V_{n+1}^c}, \mathcal{A}_{V_{n+1}})$.
 \end{theorem}
\textbf{Proof}.
From Lemma \ref{lemma_MTE_level} for each $k=1,\cdots,n$ the map $\mathcal{E}_{k,k+1}$ is a MTE with respect to the triplet
 $(\mathcal{A}_{ V_{k+1}\setminus\overset{\circ}{V}_k}, \mathcal{A}_{ V_{k+1}\setminus V_{k}},  \mathcal{A}_{\overline{V}_{k}}).$\\
 A simple induction shows that $\mathcal{A}_{V_{n+1}^c}$ is  transition expectation with respect to the given triplet.$\square$


 \section{Forward quantum Markov fields}\label{section_main}
This section will be devoted to the definition of forward quantum Markov fields on the local structure of the considered
 graph w.r.t. the considered tessellation in section \ref{section_tess}.
 Consider a product state
 \begin{equation}\label{phi0prd}
 \varphi^{(0)} = \bigotimes_{x\in V}\varphi_x^{(0)}\in \mathcal{S}(\mathcal{A}_V)
 \end{equation}
 where $\varphi_x^{(0)}$ is a state on the algebra $\mathcal{A}_x$. If $\Lambda\subseteq V$, we denote $\varphi_{\Lambda}^{(0)}:= \bigotimes_{x\in \Lambda}\varphi_x^{(0)}$.


Let $\mathcal E_{y}$ be a transition expectation w.r.t. the triplet
$(\mathcal{A}_{\{y\}\cup N_y},  \mathcal{A}_{ N^{(s)}_y}, \mathcal{A}_{\{y\}\cup N_y^{(p)}})$  for each $y\in V_{\infty}$ and
 $(\mathcal E_{n,n+1})_{1\le k\le n}$ be given by (\ref{E_vecpartialV_n}). Define for each $n\in\mathbb N$

 \begin{equation}\label{state_phi_n}
   \varphi_n = \varphi_{V_{n+1}^c}^{(0)}\circ \mathcal{E}_{n,n+1}\circ \cdots\circ \mathcal{E}_{0,1}
 \end{equation}
 From Theorem \ref{thm_Evn+1c}.  the functional $\varphi_n$ is a state on the algebra $\mathcal{A}_V$.
 \begin{definition}\label{Fwrd_QMF}{\rm
Any limit point $\varphi$ (point--wise on $\mathcal A$) of states of the form (\ref{state_phi_n}) is called a
\textbf{Forward quantum Markov fields} on $\mathcal{A}$.
The state $\varphi^{(0)}\in \mathcal{S}(\mathcal{A}_V)$ is called reference state and $(\mathcal{E}_{y})$ is called sequence of Markov transition expectations associated with
the Markov field $\varphi$.\\
If there exist two or more  limit points of  (\ref{state_phi_n}), the pair $\{\varphi^{(0)} \ , \ (\mathcal{E}_{y})_{y\in V_{\infty}}\}$ is said \textbf{to admit phase transitions}.
}\end{definition}

\begin{remark}
  In Definition \ref{Fwrd_QMF}, the Markov property is investigated on the finer structure of the considered graph. Namely, the transition expectations 
    $\mathcal{E}_{y}$  satisfies the local Markov property (\ref{MP_Plaquette}). 
     Note that  in \cite{AOM.2010}, \cite{Acc.Fid.2003} the Markov property was expressed only w.r.t. levels of the considered graphs.
 \end{remark}


\begin{definition}\label{strong-fin-lim}{\rm
A sequence $(\varphi_{n})$ of states on $\mathcal A_V$ is called
\textbf{convergent in the strongly finite sense} if, for any $a\in\mathcal A_V$,
there exists $n_a\in\mathbb N$ such that for any $n\ge n_a$,
$$
\varphi_{n}(a)=\varphi_{n_a}(a)
$$
}\end{definition}


  The following theorem is the main result of the paper
  \begin{theorem}\label{main}
  Let $G= (V,E)$ be a graph and $V_{\infty}$  be given by ( \ref{Vinfty}) and satisfy (\ref{N0=0}) and (\ref{(s)-property}).
   Let $(\varphi^0, (\mathcal{E}_y)_{y\in V_{\infty}})$ be a couple satisfying (\ref{phi0prd}) and (\ref{MP_Plaquette}).\\
 Assume that 
  \begin{equation}\label{Cpt_cd_E_y}
   \varphi^{(0)}_{  N_y^{(s)}}\circ \mathcal{E}_{y}(a_{N_y^{(p)}}\otimes
\id_{y}\otimes \id_{N_y^{(s)}}) =  \varphi^{(0)}_{N_y^{(p)}}(a_{N_y^{(p)}}); \quad  a_{N_y^{(p)}}\in \mathcal{A}_{N_y^{(p)}}
\end{equation} 
 for each $y\in V_{\infty}$. Then there exists a unique forward  quantum Markov field $\varphi$ associated with the couple $(\varphi^0, (\mathcal{E}_y)_{y\in V_{\infty}})$. Moreover, the sequence
  states $(\varphi_n)_n$ given by (\ref{state_phi_n}) converges on the strongly finite sense into a unique state $\varphi$ on $\mathcal{A}_V$.
   \end{theorem}

  \section{Proof of theorem \ref{main}}\label{proof_main}

\begin{lemma}\label{Delta_lemma}
 Let $k$ be a positive integer and $D_1, D_2,  \dots, D_k$ be subsets of a given set $X$. Denote
 $$ \Delta_0= \emptyset, \, \Delta_{j}= D_{j}\setminus (D_1\cup \dots\cup D_{j-1}),\, j=1,\,\dots,\, k$$
 then the parts $\Delta_1, \Delta_2, \dots, \Delta_k$ are pair-wise disjoint and
$$
  \bigcup_{1\le j\le k} D_j = \bigcup_{1\le j\le k}\Delta_j
$$
\end{lemma}
\begin{lemma}\label{lemm_proj}
In the notations of Theorem \ref{main}. Let $b=\bigotimes_{x\in\overleftarrow{\partial}V_n}b_x\in\mathcal A_{\overleftarrow{\partial}V_n}$. Then
\begin{equation}\label{projectivety_E}
\mathcal{E}_{n,n+1}(b\otimes \id_{\overrightarrow\partial V_n}\otimes \id_{\overleftarrow{\partial} V_{n+1}}) 
= \bigotimes_{1\leq j \leq |\overrightarrow\partial V_n|}\mathcal E_{y_{j}^{(n)}}\left(b_{\Delta_j}\otimes \id_{y_{j}^{(n)}}\otimes \id_{N^{(s)}_{y_{j}^{(n)}}}\right)
\end{equation}
where $\Delta_0=\emptyset,\quad \Delta_j= N_{y_{j}^{(n)}}^{(p)}\setminus(N_{y_{1}^{(n)}}^{(p)}\cup \cdots\cup N_{y_{j-1}^{(n)}}^{(p)}),\quad j=1,\cdots, |\overrightarrow\partial V_n|$ and $b_{\Delta_j}=\otimes_{x\in\Delta_j}b_x$.
\end{lemma}
\textbf{Proof}.
From Lemma \ref{Delta_lemma}   (with $D_j=N^{(p)}_{y_{{j}^{(n)}}}$) the sets
$\Delta_j$ are pair--wise disjoint and $\cup_{j} \Delta_j =\cup_j N_{y_{j}^{(n)}}^{(p)}
=\overleftarrow{\partial}V_n$. Without lose of generality, we assume that  $b$ is localized in the form 
$$
b= \bigotimes_{1\leq j\leq |\overrightarrow\partial V_n|}b_{\Delta_j};
\quad b_{\Delta_j}= \otimes_{x\in\Delta_j}b_x
$$
Since $\bigcup_{2\leq j\leq |\overrightarrow\partial V_n|}\Delta_j\subset (\{y_{1}^{(n)}\}\cup N_{y_{1}^{(n)}})^c$,
then,   (\ref{(s)-property}) leads to
$$
\mathcal E_{y_{1}^{(n)}}(b)=\left( \bigotimes_{2\leq j\leq |\overrightarrow\partial V_n|}b_{\Delta_j}\right)\otimes \mathcal E_{y_{1,n}}(b_{\Delta_1}\otimes\id_{y_{1,n}}\otimes \id_{N^{(s)}_{y_{1,n}}})
$$
and 
$$\mathcal E_{y_{1}^{(n)}}(b_{\Delta_1}\otimes\id_{y_{1}^{(n)}}\otimes \id_{N^{(s)}_{y_{1}{(n)}}})
\in\mathcal A_{N^{(s)}_{y_{1}^{(n)}}}\subset \mathcal A_{\{y_{j}^{(n)}\}N_{y_{j}^{(n)}}},
\quad j=2,\cdots, |\overrightarrow\partial V_n|$$
Then
$$
\mathcal{E}_{n,n+1}(b) = \mathcal E_{y_{|\overrightarrow\partial V_n|}^{(n)}}\circ\cdots\circ\mathcal E_{y_{2}^{(n)}}
\left(\bigotimes_{1\leq j\leq |\overrightarrow\partial V_n|}b_{\Delta_j}\right)\otimes\mathcal E_{y_{1}^{(n)}}(b_{\Delta_1}
\otimes\id_{y_{1}^{(n)}}\otimes \id_{N^{(s)}_{y_{1}^{(n)}}})
$$
Iterating this procedure, one gets (\ref{projectivety_E}).

\textbf{Proof}.  (proof of theorem \ref{main})

From Theorem \ref{E_Vn+1c}  and (\ref{phi0prd}) the functional $\varphi_n$ given by (\ref{state_phi_n}) is a state on the algebra $\mathcal{A}_V$
 According to the linearity of the continuity of $(\varphi_n)_n$, it is enough to show the point-wise limit on the local algebra
$\mathcal{A}_{V, loc}$.
Let $a\in\mathcal{A}_{V, loc}$. By Proposition \ref{V_n}  there exists an integer $n $ such that  $a\in\mathcal A_{V_n}$.
One has
\begin{eqnarray*}
\varphi_n(a) &=& \varphi^{(0)}_{(\overline V_n)^c}\circ \mathcal{E}_{n,n+1}\circ\dots\circ \mathcal E_{0, 1}(a),\\ 
&=& \varphi^{(0)}_{\overleftarrow{\partial} V_{n+1}}\circ \mathcal{E}_{n,n+1}(b)
\end{eqnarray*}
where $b =\mathcal E_{n-1,n} \circ\dots\circ \mathcal E_{0, 1}(a) \in \mathcal A_{\overleftarrow{\partial} V_{n+1}}$.
Up to linear extension it is enough to consider the case where $b$ is  localized in the form
$$
b =\bigotimes_{x\in \overleftarrow{\partial} V_n}b_x = \bigotimes_{x\in \overleftarrow{\partial}V_n}b_x; \quad  \quad  b_x= \id_{\mathcal A_x}\quad \forall x\in\overleftarrow{\partial}^{(p)} V_n\setminus \overleftarrow{\partial} V_n
$$
 applying above mentioned remark with $D_j= N^{(p)}_{y_{j}^{(n)}};\, 
 j=1, \,\dots,\, |\overrightarrow\partial V_n|$ and $ \Delta_{j} = 
 D_{j} \setminus (D_1\cup\dots \cup D_{j-1})\subseteq N^{(p)}_{y_{j}^{(n)}}
  ,\quad j=1, \,\dots,\, |\overrightarrow\partial V_n|$ (with $X_0=\emptyset$),
   thus again we can restrict to elements $b$ of the form:
$$
b=  \bigotimes_{j=1}^{|\overrightarrow\partial V_n|}b_{Y_j};\quad b_{Y_j}= \bigotimes_{x\in Y_j}b_x
$$
Using Lemma \ref{lemm_proj}, one obtains
\begin{eqnarray*}
\varphi_n(a) &=&  \varphi^{(0)}_{\overleftarrow{\partial} V_{n+1}}\circ \mathcal{E}_{n,n+1}(b)\\
&=& \left( \bigotimes_{j=1}^{|\overrightarrow\partial V_n|}\widehat\varphi_{N_{y_{j}^{(n)}}}\right)
 \circ\mathcal E_{y_{|\overrightarrow\partial V_n|}^{(n)}}\circ\dots\circ
  \mathcal E_{y_{1}^{(n)}}\left(\bigotimes_{j=1}^{|\overrightarrow\partial V_n|}b_{Y_j}\right)\\
&=& \prod_{1\leq j \leq |\overrightarrow\partial V_n|}\widehat\varphi_{N_{y_{j}^{(n)}}}\circ \mathcal E_{y_{j}^{(n)}}\left(b_{Y_j}\otimes \id_{y_{j}^{(n)}}
\otimes \id_{N^{(s)}_{y_{j}^{(n)}}}\right)
\end{eqnarray*}
and using condition (\ref{Cpt_cd_E_y}), one gets
$$
\varphi^{(0)}_{N_{y_{j}^{(n)}}}\circ \mathcal E_{y_{j}^{(n)}}\left(b_{Y_j}\otimes \id_{y_{j}^{(n)}}\otimes \id_{N^{(s)}_{y_{j}^{(n)}}}\right)
 = \widehat\varphi_{N^{(p)}_{y_{j}^{(n)}}}(b_{Y_j})
$$
then
\begin{eqnarray*}
\varphi_n(a) &=& \prod_{1\leq j \leq |\overrightarrow\partial V_n|} \varphi^{(0)}_{N^{(p)}_{y_{j}^{(n)}}}(b_{Y_j})\\
&=&  \varphi^{(0)}_{\overleftarrow{\partial}V_n}(b)\\
&=&   \varphi^{(0)}_{(\overline V_{n-1})^c}\circ \mathcal E_{\overrightarrow\partial V_{n-1}}\circ\dots\mathcal E_{\overrightarrow\partial V_{\infty}}(a)\\
&=& \varphi_{n-1}(a)
\end{eqnarray*}
Iterating one gets, for each $m\geq n$
$$
\varphi_m(a)= \varphi_{m-1}(a)=\cdots= \varphi_{n-1}(a)
$$
therefore the sequence $\{\varphi_n\}_n$ converges to a state
$\varphi\in \mathcal S(\mathcal{A}_V)$ for the weak-*-topology.

\section{ Annex}\label{annex}

\textbf{Proof of proposition. \ref{prop_V_n}}.
\begin{description}
 \item(i) By induction: for $n=1$ one has $V_1= \{y\}\cup N_{y}$ connected and finite (because
 the graph $G$ is locally finite). Assume that $G_n$ is connected and finite.
 One has
 $$V_{n+1}= V_n\cup \overrightarrow\partial V_n\cup \overleftarrow{\partial} V_{n+1} = V_n\cup\bigcup_{y\in \overrightarrow\partial V_n} \{y\}\cup N_y.$$
 Since the graph $G$ is locally finite and  $V_n$ is finite then $\overrightarrow\partial V_n$  and $\bigcup_{y\in \overrightarrow\partial V_n} \{y\}\cup N_y$ are also finite then $V_{n+1}$ is finite.\\
 In addition, $V_n$ is connected, every element  $y$ of $\overrightarrow\partial V_{n}$ is joined through
 one edge to some vertex $x\in V_n$.\\
 For $z\in\partial V_{n+1}$ there exits $y\in \overrightarrow\partial V_n$ such that $z\in N_y$
 therefore $z\sim y \sim x$ for some $x\in V_n$. Therefore, $G_{n+1}=(V_{n+1}, E_{n+1})$ is also connected.
 \item{(ii)} Let $x\in V$, denoting $d = d(x,y_1)$ and let $x_0=y_1\sim x_1\sim \dots\sim x_d =x$
 one has
 $$x_1\sim y_1\Rightarrow x_1\in N_{y}\subset V_1, \,  \cdots, \,
 x=x_d\sim x_{d-1}\in V_{d-1}\Rightarrow x_d\in \overline V_{d-1}\subset V_d$$
     then $V\subset \bigcup_{n}V_n$, the second inclusion is obvious.\\
By construction the sequence $(V_n)_n$ is increasing and it absorbs all finite sets of $V$. We conclude that  
$V = \bigcup_{n}V_n$ and $E= \bigcup E_n$.  $\square$
\end{description}

\textbf{Proof of Proposition. \ref{partition}}
\begin{enumerate}
\item[(i)] It's clear that $\bigcup_{y\in \overrightarrow\partial V_n }N^{(s)}_{y}\subset \overleftarrow{\partial} V_{n+1}$.

Conversely, by construction $V_{n+1}= V_n\cup\overrightarrow\partial V_n\cup\overleftarrow{\partial} V_{n+1}$, the plaquette at each element of $V_n\cup \overrightarrow\partial V_n$ is
included in $V_{n+1}$ thus
$$\overleftarrow{\partial} V_{n+1}\subset V_{n+1}\setminus (V_n\cup \overrightarrow\partial V_n) = \bigcup_{y\in\overrightarrow\partial V_n}N_y$$

then $$\overleftarrow{\partial} V_{n+1} = \bigcup_{y\in \overrightarrow\partial V_n}N_y\cap V_{n+1}= \bigcup_{y\in \overrightarrow\partial V_n}N^{(s)}_{y}$$

\item[(ii)] By definition of $\overleftarrow{\partial} V_n$ one has
 $$(\forall x\in\overleftarrow{\partial}{V}_n, \exists y\in V_n^c; x\sim y) \Leftrightarrow \forall x\in\overleftarrow{\partial}{V}_n, \exists y\in \overrightarrow\partial V_n; x\in N_y $$
 and since  $\overleftarrow{\partial} V_n\cap\overleftarrow{\partial} V_{n+1}=\emptyset$ and $N^{(s)}_y\subset \overleftarrow{\partial} V_{n+1},$ one get $x\in N^{(p)}_y$ therefore, $\overleftarrow{\partial} V_n\subset \bigcup_{y\in\overrightarrow \partial V_n}N_y^{(p)}$.
\end{enumerate}

 \textbf{Proof of Proposition. \ref{tree_S_property} }.
  Let $G= (V,E)$ be a tree occupied  with the sets  $\{ V_{0,n}, n\in\mathbf N\}$  given by ( \ref{v0,n+1}). \\
  By Proposition \ref{prop_V_n}  the graph $G_n =(V_n, E_n) $ is connected. 
  and since each vertex from $\overrightarrow\partial V_n$ is related to some vertex from
   $V_n$ then $\overline V_n$ is connected, in particular every edge from $\overline V_n$ is joined
    to the root $o$ through an edge path. Let $y$ and $z$ be two disjoint vertices of $\overrightarrow\partial V_n$, 
    consider two edge-paths $\gamma_{y, y_1}: u_1=y\sim u_2\sim  \dots \sim u_k =y_1$ and $\gamma_{y_1, z}: v_1=y_1\sim v_2 \sim \dots \sim v_m= z$ 
    joining respectively $y$ with $y_0$ and $y_0$ with $z$.  If $x\in N^{(s)}\cap N^{(p)}$ 
   then
 $$\gamma: x\sim (y=u_1)\sim\dots \sim (u_k= o = v_1)\sim v_2\sim \dots \sim (v_m= z)\sim x$$
  is a cycle on the tree   $G$ is a tree. Then  $N^{(s)}\cap N^{(s)}=\emptyset$ and  the tree $G$ enjoys the  (\ref{(s)-property}).


\end{document}